\documentstyle[twoside,psfig,11pt]{article}
\newcommand{\titsep}{\vspace*{0.20cm}}
\newcommand{\secsep}{\vspace*{0.50cm}}

\def\spose#1{\hbox to 0pt{#1\hss}}
\def\lta{\mathrel{\spose{\lower 3pt\hbox{$\mathchar"218$}}
     \raise 2.0pt\hbox{$\mathchar"13C$}}}
\def\gta{\mathrel{\spose{\lower 3pt\hbox{$\mathchar"218$}}
     \raise 2.0pt\hbox{$\mathchar"13E$}}}

\def\etal{{\it et al.\ }}

\begin{document}
\emergencystretch=20pt
\begin{center}
{\LARGE \bf Tidal Dwarf Galaxies:}

\bigskip
{\LARGE \bf  Their Present State and Future Evolution}
\vspace*{0.3cm}

\markboth{Fritze - v. Alvensleben \& Duc}{Tidal Dwarf Galaxies}

{\large \bf U. Fritze - v. Alvensleben$^1$ and P.-A. Duc$^2$}
\vspace*{0.3cm} \\
 {$^1$Universit\"ats-Sternwarte G\"ottingen: $^2$ESO Garching}

\vspace*{0.85cm}

\setlength{\parindent}{0.5cm}
\hspace*{-0.6cm} \begin{minipage}{14.0cm}
Evolutionary synthesis models for Tidal Dwarf Galaxies (TDGs) are presented
that allow to have varying proportions of young stars formed in the
merger-induced starburst and of stars from the merging spirals' disks.
Comparing model grids with observational data (see e.g. P.-A. Duc this
conference for a review) we try to identify the present evolutionary state
of TDGs. The influence of their specific metallicities as well as of the
gaseous emission of actively star forming TDGs on their luminosity and colour
evolution are studied. 
\end{minipage}
\end{center} \secsep


{\noindent \large \bf 1. Motivation and Method} \titsep \\
P.-A. Duc (this volume) presents our present state of observational 
knowledge on Tidal Dwarf Galaxies ({\bf TDGs}). The aim of the present 
investigation is to understand the past and present stages of evolution of 
TDGs in order to be able to restrict possible future evolutionary paths of 
these objects. While as early as 1956 Zwicky had already considered the possibility that self-gravitating objects 
in tidal tails might acquire dynamical independence and contribute to the population 
of dwarf galaxies, by today there are basically two alternative scenarii for the 
formation of TDGs:

\begin{itemize}
\item Stellar-dynamical models reveal local concentrations of {\bf stars}  
along stellar tidal tails torn out from the disk of an interacting spiral 
(Barnes \& Hernquist 1992). Gas, if present, 
may then fall into the potential well defined by disk stars. 
\item Hydro-dynamical models show local instabilities of {\bf gas} along gaseous tidal 
tails that give rise to 
Super-Giant Molecular Clouds, which then may ignite a Burst of Star 
Formation (Elmegreen \etal 1993). Some stars, if present, might then fall into the 
potantial defined by the gaseous component. 
\end{itemize}

\noindent
The method we use is chemical and spectrophotometric evolutionary synthesis, 
i.e. our modelling starts from a gas cloud, specifies two basic parameters -- 
the Star Formation Rate in its time evolution ({\bf SFR}(t)) and the IMF -- and then follows the evolution 
of ISM abundances {\bf and} of spectrophotometric properties of the stellar 
population including gaseous line and continuum emission for various metallicities. 
Our models, of course, have to rely on various pieces of input physics. Here, in particular, we use 
new Geneva stellar evolutionary tracks for various metallicities  Z $=10^{-3},~4 \cdot 10^{-3},~8 \cdot 
10^{-3},~2 \cdot 10^{-2},~4 \cdot 10^{-2}$ 
(Schaller \etal 1992, Schaerer \etal 1993a, b, Charbonnel 
\etal 1993, 1996), 
Lyman continuum photons N$_{\rm Lyc}$ from Schaerer \& de Koter (1997), and  
emission line ratios of some 30 strong lines relative to H$_{\rm \beta}$ 
from photoionisation models (Stasi\'nska 1984) for Z$_{\odot}$, and 
from HII region observations (Izotov \etal 
1994) for ${\rm Z < Z_{\odot}}$, and also include gaseous continuum emission. 

\smallskip \noindent
Basic parameters for our modelling are the IMF, which we take from Scalo over the mass
range from 0.1 $\dots$ 85 M$_{\odot}$ -- and assume it to be identical for the 
interacting galaxies, the
starburst, and the TDGs for simplicity -- and the SFR ${\rm \Psi(t)}$, which, for spiral
galaxies we take to linearly depend on the gas-to-total mass ratio 
$\Psi \sim {\rm M_{Gas}/M_{tot}}$ with characteristic 
times for SF ${\rm t_{\ast}} = 3,~ 10,~ 16$ Gyr for Sb, Sc, Sd spirals. 
For these SF histories our models were shown to give agreement both with the chemical
properties of spiral galaxies of various types as seen in HII region abundance 
observations as well as in the redshift evolution of Damped Ly$\alpha$ Absorber abundances 
(Fritze - v. A. \etal 1997) {\bf and} with the spectral properties of nearby galaxy
templates as well as with the photometric properties from galaxy redshift surveys (e.g. Lindner \etal 1996).

\secsep \vspace*{0.2cm}
{\noindent \large \bf 2. Metallicities of TDGs}  \titsep \\
From our modelling of spiral-spiral mergers -- including the starbursts 
triggered by the interaction process in gas-rich systems -- we predicted 
metallicities for stars and star clusters forming in the burst on the basis of the spirals 
galaxy ISM abundances. This metallicity prediction, of course, also applies to young stellar 
populations of TDGs. For gas-rich spirals of ages 8 -- 12 Gyr, stars are expected to 
form with metallicities in the range [O/H] $= -0.7 ~ \dots ~ 0.0$ (Fritze -- v. Alvensleben \& Gerhard 1994a,b). 
HII region abundances of TDGs fall well within this range: 
[O/H] (TDGs) $= -0.63 ~\dots ~ -0.33 $ and are
significantly higher, on average, than values derived from the L--Z relation 
for dwarfs (cf. Fig. 3, Duc \etal, {\sl this volume}).

\secsep \vspace*{0.2cm}

{\noindent \large \bf 3. Model Grid} \titsep \\
Our models for nearby TDGs include two ingredients in various proportions 
in accordance with the two possible formation scenarii: 
\begin{itemize}
\item a composite stellar population of age $\sim 12$ Gyr from the progenitor disk
(Sb, Sc, Sd) and 
\item a starburst of given strength and duration. 
\end{itemize}
In the Barnes \etal formation scenario we expect a TDG to consist of a large contribution 
of disk stars, evolving passively since the tidal tail has been extracted from the spiral, and 
some {\sl a priori} unknown contribution of young stars forming in a starburst from the gas trapped 
into the stellar potential. In the Elmegreen \etal scenario a dominant young burst population is 
expected. 
In our attempt to understand the presently available sample of TDGs, we calculate a grid of models 
covering various progenitor populations (Sb, Sc, Sd, all assumed to be between 8 and 12 Gyr old), various burst 
strengths and durations, and we also explore the influence of changes in the metallicities with respect to 
the predicted one. 

\medskip\noindent
The aim of the present investigation is to 
age-date the dominant stellar populations using the appropriate metallicity, to 
determine relative contributions from parent disk and young burst stars, to estimate the burst duration 
and to use all this information 
to predict the future luminosity and colour evolution. 
Gas reservoirs observed in HI on several TDGs are usually large enough to fuel SF at the 
present rate for Gyrs. Dynamical effects, as e.g. a possible fall-back of some TDGs onto the 
merger remnants (cf. Hibbard \& Mihos 1995) or disruption by the parent galaxy, a group or a 
starburst-driven wind, are alltogether {\bf not} included in our models. 

\secsep \vspace*{0.2cm}
{\noindent \large \bf 4. First Results} \titsep \\
{\noindent \bf 4.1. Gaseous Emission}

\smallskip\noindent
Fig. 1. shows the enormous importance of the gaseous emission for the broad band 
colours UBVRIJHK during the active burst phase, confirming 
Kr\"uger \etal's  (1995) results obtained for starbursts in Blue Compact Dwarf galaxies. 
Decomposition of the total gaseous contribution in terms of 
line and continuum emission shows that while line emission is dominant in the optical, 
continuum emission dominates in the NIR.

\begin{figure}[h]
\centerline{\psfig{file=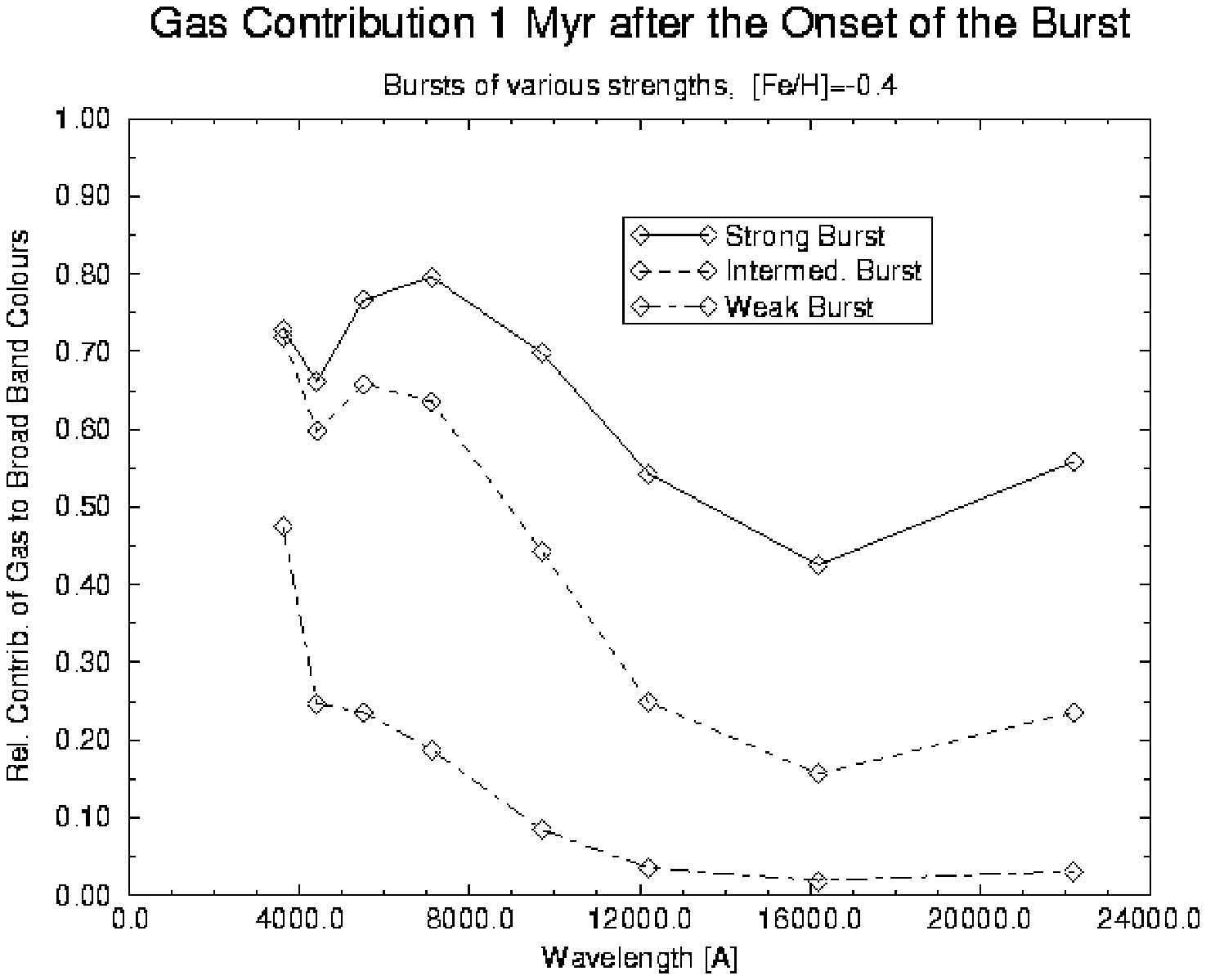,width=8cm,angle=180}
\psfig{file=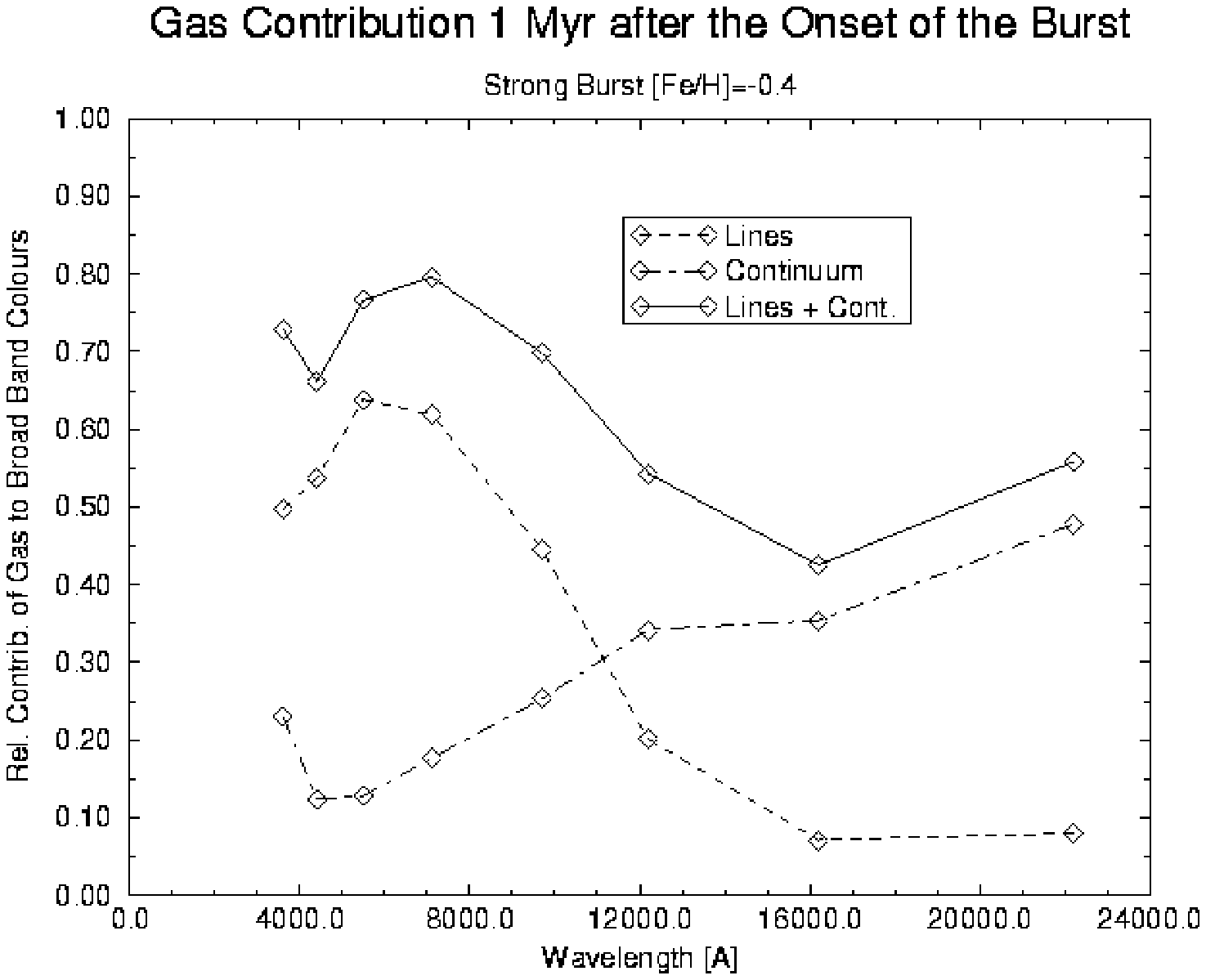,width=8cm,angle=180}}
\caption{a) Relative contribution of gaseous emission to broad band colours UBVRIJHK 1 Myr after the onset of 
bursts of various strengths. 
b) Decomposition of the total gaseous emission contribution for a strong burst (Fig.1 a) into contributions 
from lines and continuum.}
\end{figure}

\noindent
The relative contributions of the young population formed {\sl in situ} from a gas condensation in the 
tidal tail and the 
``old'' (=composite in age and metallcity) population extracted from a spiral disk for our strong, intermediate, 
and weak burst models to the integrated light in B, V, and K bands, as well as to the total stellar mass S 
are displayed in Tab.1. 
 
\begin{table}[h] 
  \begin{center} 
  \begin{tabular}{|c|c|c|c|c|} 
   \hline
       Burst & ${\rm L_B^{young}}/{\rm L_B^{old}}$ & ${\rm L_V^{young}}/
       {\rm L_V^{old}}$ & ${\rm L_K^{young}}/{\rm L_K^{old}}$ & 
       ${\rm S^{young}}/{\rm S^{old}}$\\
   \hline
       Strong &  26 & 20 & 2.7 & 0.4\\ 
       Intermediate & 7 & 5 & 1.4 & 0.1 \\ 
       Weak & 1.6 & 1.4 & 1.04 & 0.01\\ 
   \hline
  \end{tabular}
  \caption{Relative contributions to luminosities in B,V, and K bands and to stellar 
  mass S of young burst and ``old'' disk stars.}
  \end{center} 
\end{table} 

\noindent
It should be noted that while our starburst models give very good agreement with the standard 
formula for the transformation of H$_{\alpha}$ luminosity into SFR (Hunter \& Gallagher 1986) for SFRs 
up to a few M$_{\odot}$yr$^{-1}$, they show that for SFRs ${\rm > 10 ~M_{\odot}yr^{-1}}$ the standard 
SFR(L$_{\rm H_{\alpha}}$) strongly underestimates the true SFR (by a factor $\sim 3.6$ for 
SFRs $\gta 50$ M$_{\odot}$yr$^{-1}$).  

\medskip
{\noindent \bf 4.2. Colour Evolution}

\smallskip\noindent
Fig. 2 gives the colour evolution through and after bursts of various 
strengths on top of a 12 Gyr Sc population (u.g.). Model curves are for 
[Fe/H] $= -$0.4. 
Starting from the colour of the u.g. at 12 Gyr, model galaxies evolve clockwise 
along the curves. Evolution to the bluest point takes $\sim$ 7 Myr, from there back to 
the u.g. colours takes 16 to 30 Myr, and from there to the end of the graph 
$\sim$ 2.6 Gyr. Pure burst models without any underlying population from the spiral disk 
would start in the upper left corner of the diagrams and evolve toward the lower left on the same 
passive evolutionary reddening path as the strong burst models. 
TDGs from Duc's compilation are plotted twice: 
$\Diamond$ : corrected for the total extinction A$_{\rm B}$ as derived from the Balmer decrement, 
$\Diamond$ : corrected for  Galactic extinction only. 
While the extinction from the Balmer decrement might somehow overestimate the extinction on 
integrated colours, the  Galactic extinction gives a lower limit. 
Typical errors for the colours are of order $\sim$ 0.2 mag. 

\smallskip\noindent
It is seen from Figs 2. that the bulk of the TDGs from Duc's sample seem to feature quite strong starbursts 
on top of a composite stellar population of disk stars. Due to the uncertainties in the reddening 
correction it is not clear, though, if within the present sample of TDGs, there really are pure bursts 
without underlying component or purely ``old'', passively evolving stellar subsystems from spiral disks. 

\begin{figure}[h]
\centerline{\psfig{file=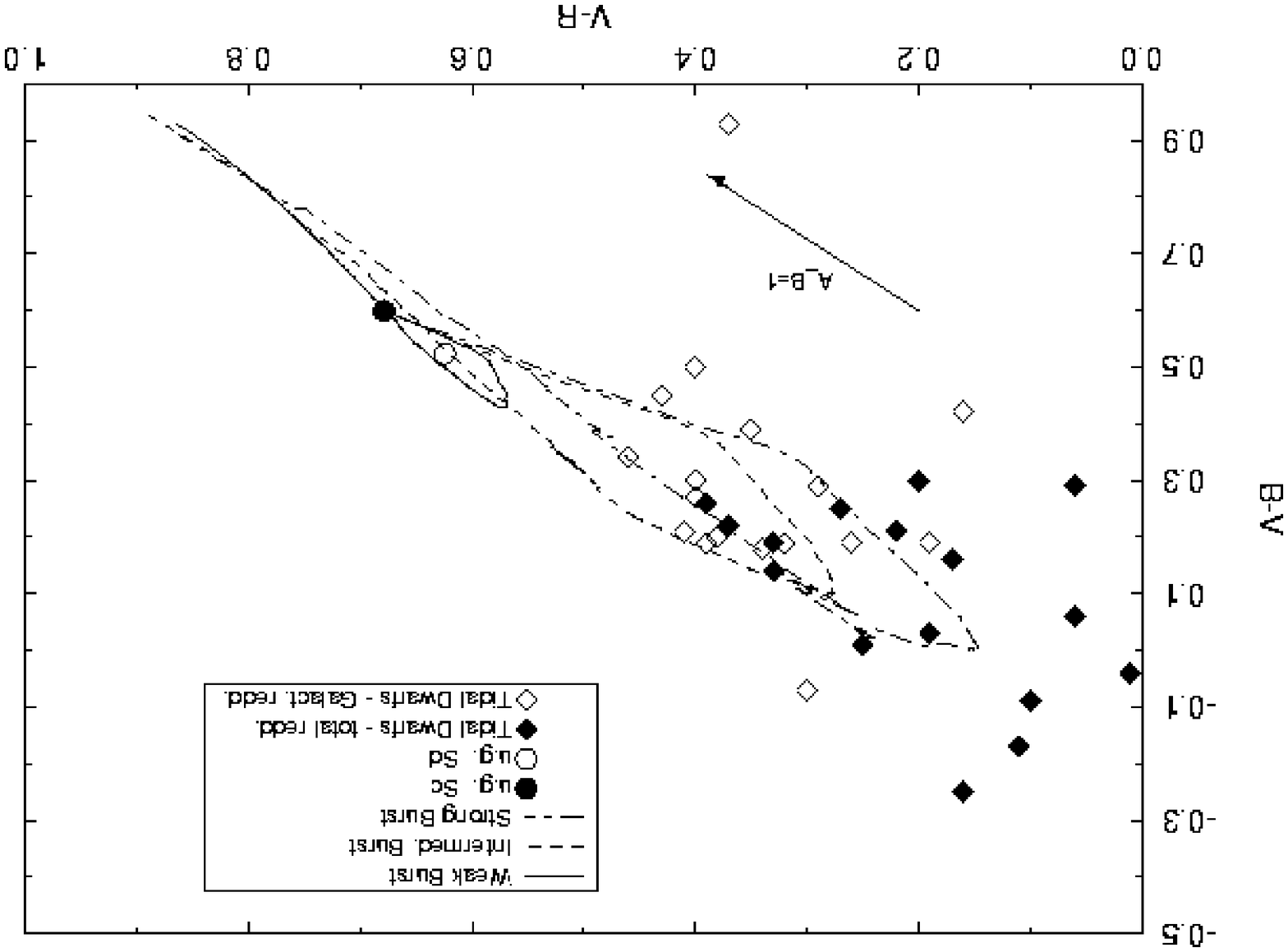,width=8cm,angle=180}
\psfig{file=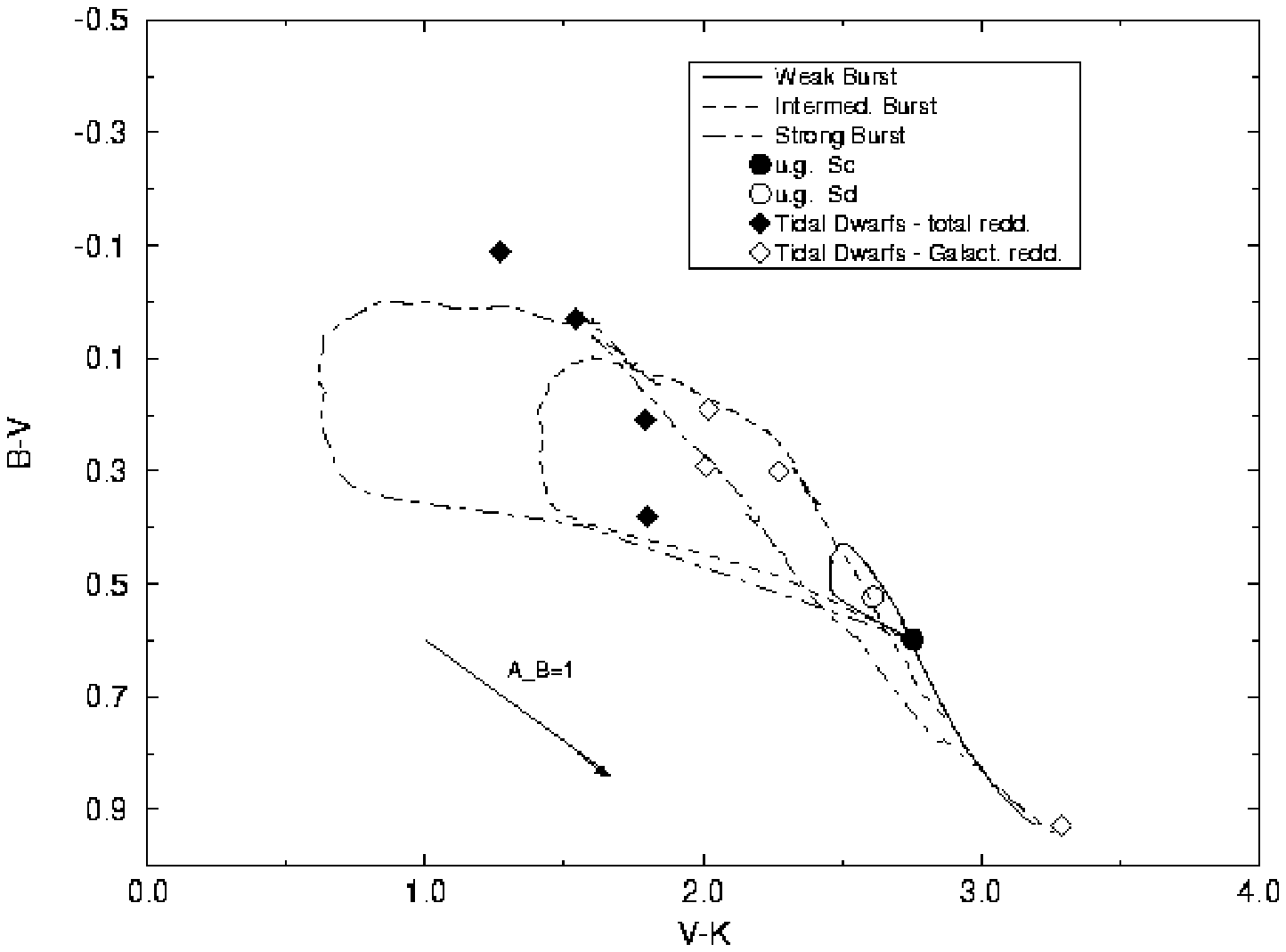,width=8cm,angle=180}}
\caption{Two colour diagrams for $(B-V)$ vs. $(V-R)$ {\bf a)} and $(B-V)$ vs. $(V-K)$ {\bf b)}, 
respectively.}
\end{figure}

\newpage
{\noindent \bf 4.3. Luminosity Evolution}

\smallskip\noindent
The stronger the light contribution of the young burst star 
population, the stronger the fading. 
For TDGs with a strong burst population the fading in B 
within $\sim$ 1 Gyr may easily amount to 2 -- 3 mag, slightly depending on [Fe/H], while in K the 
brightening as well as the fading during the 1$^{\rm st}$ Gyr even of a strong burst hardly exceed 1 mag. 
A correct estimate of the ``old'' star population is crucial for a reliable fading prediction. 

\medskip
{\noindent \bf 4.4. A105S: a first example}

\smallskip\noindent
Emission line spectroscopy gives a metallicity [O/H] $= -0.6$, colours from B through K indicate a strong burst 
of short duration ${\rm t_{\ast} \sim 10^6}$ yr and a burst age of $(2 - 6) \cdot 10^7$ yr. The stellar 
mass we estimate using the M/L$_{\rm K}$ at this age from our model is ${\rm \sim 4 \cdot 10^8~M_{\odot}}$. 
Together with the observed ${\rm M(HI) \sim 5 \cdot 10^8~M_{\odot}}$ it reasonably agrees with the dynamical 
mass ${\rm M_{dyn} \sim 1 \cdot 10^9~M_{\odot}}$ obtained from the rotation curve. While the light contribution 
from the ``old'' disk star population accounts for 44 \% of the total ${\rm L_K}$ it only make up 4 \% of 
the present ${\rm L_B}$. The mass of the old population derived from ${\rm L_K}$ is ${\rm \sim 3 \cdot 
10^8~M_{\odot}}$. 
The stellar mass of A105S thus is by far dominated by the ''old'' population. This has important implications for 
both the dynamical and the photometric evolution of this TDG. 

\secsep \vspace*{0.2cm}
{\noindent \bf 5. Outlook} \titsep \\
Next steps will obviously be to study burst strengths and fading for a sample of TDGs in order to define the 
range of burst strengths and durations realised in TDGs, in particular to assess the question if there are 
purely young starburst TDGs and purely ``old'' passively evolving disk star condensations ($\rightarrow~~$ 
Diploma thesis P. Weilbacher).  
Only then can our speculation about a possible contribution 
of TDGs to the Faint Blue Galaxy excess be put on quantitative grounds. 

\medskip
{\noindent  \bf Acknowledgement}\\
It is a pleasure to thank the organisers for this very inspiring workshop, the Physikzentrum and the DARA 
(50 OR 9407~6) for financial support.

\secsep \vspace*{0.2cm}

{\noindent \large \bf References} \titsep \vspace*{-0.2cm}
\begin{description}
\itemsep=0pt \parsep=0pt \parskip=0pt \labelsep=0pt

\item Barnes, J.~E., Hernquist, L., 1992, Nat. {\bf {360}}, 715
\item Charbonnel,C., Meynet,G., Maeder,A., Schaller,G., Schaerer,D., 1993, A\&A Suppl. {\bf {101}}, 415
\item Charbonnel, C., Meynet, G., Maeder, A., Schaerer, D., 1996, A\&A Suppl. {\bf {115}},~339
\item Duc P.-A., Brinks E., Wink J.~E. and Mirabel I.~F., 1997, A\&A 326, 537
\item Duc, P.-A., Mirabel, I. F., 1994, A\&A {\bf {289}}, 83
\item Elmegreen, B.~G., Kaufman, M., Thomasson, M., 1993, ApJ 
        {\bf {412}}, 90
\item Fritze - v. Alvensleben, U., Gerhard, O.~E., 1994a, A\&A 
   	{\bf {285}}, 751
\item Fritze - v. Alvensleben, U., Gerhard, O.~E., 1994b, A\&A 
  	 {\bf {285}}, 775
\item Fritze - v. Alvensleben, U., Lindner, U., Fricke, K.~J., 1998, IAU Symp. 187 {\sl in press}
\item Hibbard J.~E. and Mihos J.~C. 1995, AJ 110, 140
\item Hunter, D.~A., Gallagher, J.~S., 1986, PASP {\bf 98}, 5
\item Izotov, Y.~I., Thuan, T.~X., Lipovetsky, V.~A., 1994, ApJ 
        {\bf {435}}, 647
\item Kr\"uger, H., Fritze - v. Alvensleben, U., Loose, H.-H., 1995, A\&A 
   	{\bf {303}}, 41
\item Lindner, U., Fritze - v. Alvensleben, Fricke, K.~J., 1996, A\&A {\bf 316}, 132
\item Schaerer, D., de Koter, A., 1997, A\&A {\bf 322}, 598
\item Schaerer, D., Meynet, G., Maeder, A., Schaller, G., 1993, A\&A Suppl. 
        {\bf {98}}, 523
\item Schaerer,~D., Charbonnel,~C., Meynet,~G., Maeder,~A., Schaller,~G., 1993, A\&A Suppl.~{\bf {102}},~339
\item Schaller, G., Schaerer, D., Meynet, G., Maeder, A., 1992, A\&A Suppl. 
        {\bf {96}}, 269
\item Stasi\'nska, G., 1984, A\&A Suppl. {\bf {55}}, 15
\item Zwicky, F., 1956, Ergebn. der Exakten Naturwiss., {\bf 29}, 344
\end{description}
\end{document}